THE EUROPEAN
PHYSICAL JOURNAL C

Regular Article - Theoretical Physics

# Observing $t\bar{t}Z$ spin correlations at the LHC

Baptiste Ravina[a], Ethan Simpson[b], James Howarth[c]

School of Physics and Astronomy, University of Glasgow, Glasgow G12 8QQ, UK

**Abstract** Spin correlations in the production of top-antitop quark ($t\bar{t}$) pairs at the Large Hadron Collider (LHC) are an experimentally verified prediction of the Standard Model. In this paper, we compute the full spin density matrix for $t\bar{t}Z$ production at next-to-leading order precision in QCD, for center-of-mass energies of 13 and 14 TeV. We find that the additional emission of a $Z$ boson leads to significantly different spin correlations with respect to the $t\bar{t}$ case, and induces small longitudinal polarisations of the top quarks. We further propose an analysis strategy that could lead to the observation of spin correlations in $t\bar{t}Z$ events at the end of Run 3 of the LHC, or possibly earlier by combining the ATLAS and CMS datasets. In addition, we show that the pure angular information contained in the spin density matrix provides novel constraints on the dimension-6 effective field theory operators relevant to the $t$-$Z$ interaction, without any reference to the total production rates.

## Contents

[a] e-mail: baptiste.ravina@cern.ch (corresponding author)
[b] e-mail: ethan.simpson@cern.ch
[c] e-mail: jhowarth@cern.ch

## 1 Introduction

The top quark is the only quark in the standard model (SM) to decay on a timescale shorter than that of strong interactions. As a consequence, the top quark can be treated as a bare quark and its fundamental quantum properties probed through sensitive observables built from its decay products. In high energy particle colliders, tops are most frequently produced in pairs of top-antitop quarks ($t\bar{t}$) and it is therefore possible to measure the correlation between the spins of the $t$ and $\bar{t}$, as well as their individual polarisations. At the LHC, spin correlation in $t\bar{t}$ events were first observed by the ATLAS experiment in Ref. [1], and both the ATLAS and CMS experiments have presented measurements of $t\bar{t}$ spin correlations and top polarisations in multiple final states [2–8]. A recent ATLAS analysis [5] originally reported a deviation in their measurement of spin correlations, based on a SM prediction at NLO accuracy in QCD. This result has been extensively discussed, and compared to dedicated calculations both at NLO QCD including electroweak effects [9,10], and at NNLO QCD [11,12].

With the current level of experimental precision, such measurements are becoming powerful probes of the SM, particularly relying on the excellent identification and reconstruction of charged leptons at the ATLAS and CMS experiments. Within the framework of the Standard Model Effective Field Theory (SMEFT), which provides a model-independent and systematic categorisation of any effects from possible new physics at higher energies, these precise angular measurements have been shown to provide additional sensitivity to certain SMEFT operators related to $t\bar{t}$ production [4,13]. In this paper, we turn our attention to the case of top-antitop-quark pair production in association with a $Z$ boson ($t\bar{t}Z$) and show that the top-$Z$ coupling induces significantly different spin correlations.

The most recent results of $t\bar{t}Z$ cross section measurements by the ATLAS collaboration at 139.1 fb$^{-1}$ [14] and the CMS collaboration at 77.5 fb$^{-1}$ [15] show a good experimental agreement of the inclusive rate with the NLO QCD and electroweak corrections calculation of Ref. [16] (based on Ref. [17]), as well as the more recent state-of-the-art predictions at NLO QCD + NNLL accuracy and matched to the complete NLO results [18,19]. These two analyses also provide the first $t\bar{t}Z$ differential cross section measurements in a number of key observables, including kinematics of the $Z$ boson. However, none of these observables can be directly re-interpreted as a measure of the amount of correlation between the spins of the top quarks after emission of the $Z$ boson. The CMS paper [15] further presents a re-interpretation of its results in the context of the SMEFT, drawing exclusion power from both the rate and shape information. We shall argue here that angular distributions built from the decay products of the top quarks provide complementary sensitivity to existing approaches and, crucially, without relying on any binned cross section dependence. The use of multiple additional normalised differential distributions allows us to extract more information from a given measurement while keeping statistical correlations low, and could eventually help isolate the contributions of individual SMEFT operators.

This paper is structured as follows: in Sect. 2, we briefly review the spin density matrix formalism, highlighting the connection between its components and specific angular observables. We then provide predictions at NLO QCD precision, assuming the SM. In Sect. 3, we propose an analysis strategy to perform a template fit to selected angular distributions and observe spin correlations in $t\bar{t}Z$ events. We estimate the current experimental sensitivity using the full LHC Run 2 data, and extrapolate it according to several scenarios, including higher center-of-mass energies and integrated luminosities. Finally, we show in Sect. 4 that departures from our SM predictions can be reframed in the SMEFT in terms of the higher-dimension operators involved in the anomalous $t$-$Z$ coupling. We conclude in Sect. 5.

## 2 Spin density matrix

Following the formalism of Ref. [20], we recall that hadronic $t\bar{t}$ production in the narrow-width approximation depends on a spin density matrix $R$, decoupled from the decay of the top quarks (which is hereafter assumed to proceed entirely via $t \to Wb$). In the spin spaces of $t$ and $\bar{t}$ (indexed as + and − respectively), the matrix $R$ is

$$R \propto A \mathbb{1} \otimes \mathbb{1} + B_i^+ \sigma^i \otimes \mathbb{1} + B_i^- \mathbb{1} \otimes \sigma^i + C_{ij} \sigma^i \otimes \sigma^j. \quad (1)$$

While the function $A$ encodes the $t\bar{t}$ cross section and parton kinematics, the 3-vectors $\mathbf{B}^\pm$ represent the top polarisations and the $3 \times 3$ matrix $C$ the correlation between the spins of the top and anti-top quarks. To define these three axes, we adopt the basis of Ref. [20]: taking $\hat{\mathbf{k}}$ as the top-quark direction in the $t\bar{t}$ centre-of-mass frame and $\hat{\mathbf{p}}$ as the direction of one of the incoming proton beams in the laboratory frame, we have

$$y = \hat{\mathbf{p}} \cdot \hat{\mathbf{k}}, \quad r = \sqrt{1 - y^2},$$
$$\hat{\mathbf{r}} = \frac{\text{sign}(y)}{r} \left( \hat{\mathbf{p}} - y\hat{\mathbf{k}} \right), \quad \hat{\mathbf{n}} = \frac{\text{sign}(y)}{r} \left( \hat{\mathbf{p}} \times \hat{\mathbf{k}} \right), \quad (2)$$

and $\{\hat{\mathbf{r}}, \hat{\mathbf{k}}, \hat{\mathbf{n}}\}$ is a right-handed orthonormal basis. The extra factor of $\text{sign}(y)$ in the definitions of $\hat{\mathbf{r}}$ and $\hat{\mathbf{n}}$ breaks the Bose symmetry of the $gg$ initial state by identifying a forward direction and ensures non-zero values of the relevant spin density matrix elements.

The polarisation vector $\mathbf{B}^\pm$ can now be expressed in terms of three functions $b_i^\pm$, which depend on the center-of-mass energy $\sqrt{s}$ of the proton-proton collision and the top-quark scattering angle $y$:

$$\mathbf{B}^\pm = b_r^\pm \hat{\mathbf{r}} + b_k^\pm \hat{\mathbf{k}} + b_n^\pm \hat{\mathbf{n}}. \quad (3)$$

At leading order and in the absence of extra quarks in the final state, CP-invariance implies that $b_i^+ = b_i^-$. The $b_r^\pm$ and $b_k^\pm$ components induce a longitudinal polarisation of the top and anti-top quarks, while $b_n^\pm$ leads to a transverse polarisation; however, under a T-transformation $b_n^\pm(y) \to -b_n^\pm(y)$, such that $b_n^\pm$ must vanish at tree-level in QCD.

Similarly, the spin correlation matrix $C$ can be expanded in terms of the basis (2) as

$$\begin{aligned}
C &= c_{rr}\hat{\mathbf{r}}\hat{\mathbf{r}}^\top + c_{kk}\hat{\mathbf{k}}\hat{\mathbf{k}}^\top + c_{nn}\hat{\mathbf{n}}\hat{\mathbf{n}}^\top \\
&+ c_{rk}\left(\hat{\mathbf{r}}\hat{\mathbf{k}}^\top + \hat{\mathbf{k}}\hat{\mathbf{r}}^\top\right) + c_{kn}\left(\hat{\mathbf{k}}\hat{\mathbf{n}}^\top + \hat{\mathbf{n}}\hat{\mathbf{k}}^\top\right) \\
&+ c_{rn}\left(\hat{\mathbf{r}}\hat{\mathbf{n}}^\top + \hat{\mathbf{n}}\hat{\mathbf{r}}^\top\right) \\
&+ c_r\left(\hat{\mathbf{k}}\hat{\mathbf{n}}^\top - \hat{\mathbf{n}}\hat{\mathbf{k}}^\top\right) + c_k\left(\hat{\mathbf{n}}\hat{\mathbf{r}}^\top - \hat{\mathbf{r}}\hat{\mathbf{n}}^\top\right) \quad (4) \\
&+ c_n\left(\hat{\mathbf{r}}\hat{\mathbf{k}}^\top - \hat{\mathbf{k}}\hat{\mathbf{r}}^\top\right), \quad (5)
\end{aligned}$$

where the components $c_i$ and $c_{ij}$ are again functions of $\sqrt{s}$ and $y$. The three antisymmetric cross-correlations ($c_r$, $c_k$ and $c_n$) are sourced by CP-violation and therefore suppressed in the SM, while two of the symmetric ones ($c_{kn}$ and $c_{rn}$) only appear in mixed QCD-weak 1-loop corrections and are also very small. On the other hand, $c_{rk}$ and the three diagonal correlations ($c_{rr}$, $c_{kk}$ and $c_{nn}$) are all C-, CP- and CPT-even, and therefore significantly non-zero at LO in QCD[1].

2.1 Relation to angular observables

Given two unit vectors $\hat{\boldsymbol{\psi}}_+$ and $\hat{\boldsymbol{\chi}}_-$, corresponding to the direction of flight of two visible decay products of the top and anti-top quarks in their respective rest frames, we can express the normalised four-fold angular distributions in terms of the polarisation vectors $\mathbf{B}^\pm$ and the spin correlation matrix $C$ as [13]:

$$\frac{1}{\sigma}\frac{d\sigma}{d\Omega_+ d\Omega_-} = \frac{1}{(4\pi)^2}\left(1 + \kappa_\psi \mathbf{B}^+ \cdot \hat{\boldsymbol{\psi}}_+ + \kappa_\chi \mathbf{B}^- \cdot \hat{\boldsymbol{\chi}}_+ - \kappa_\psi \kappa_\chi \hat{\boldsymbol{\psi}}_+ \cdot C \cdot \hat{\boldsymbol{\chi}}_-\right), \quad (6)$$

where $d\Omega_\pm = d\cos\theta_\pm d\phi$ defines a solid angle in polar coordinates. The pre-factors $\kappa$ are known as "spin analysing powers" and provide a measure of the extent to which the spin information is diluted through the $V-A$ structure of the top quark decay [21]. The spin analysing power of charged leptons is $\kappa_\ell = 1$ at tree-level (and receives NLO QCD corrections below the per-mill level). While for up-type and $b$-quarks it is much smaller, we note that $\kappa_d \sim 0.97$ such that one might be tempted to consider the light down-type jets from a hadronic $W$ boson decay as well. One strategy proposed before in Ref. [21] is to consider the dominant $W^+ \to c\bar{s}$ hadronic mode and apply a $c$-tagger to identify the $\bar{s}$-jet. Due to the current low efficiency of $c$-taggers (complicated by the presence of additional $b$-jets in the event), no experimental $t\bar{t}$ analysis has yet put this approach to fruition[2]; meanwhile the $t\bar{t}Z$ production rates, almost three orders of magnitude smaller, clearly do not allow for the luxury of a $W^+ \to c\bar{s}$ selection if any hope of statistical precision is to be retained.

---

[1] As is of course the $A$ function of Eq. (1).

We therefore take $\hat{\boldsymbol{\psi}}_+ = \hat{\boldsymbol{\ell}}_+$ and $\hat{\boldsymbol{\chi}}_- = \hat{\boldsymbol{\ell}}_-$ (with $\ell = e, \mu, \tau$). For a choice of reference axes $\hat{\mathbf{a}}$ and $\hat{\mathbf{b}}$ amongst $\{\hat{\mathbf{r}}, \hat{\mathbf{k}}, \hat{\mathbf{n}}\}$, we define:

$$\cos\theta_+ = \hat{\boldsymbol{\ell}}_+ \cdot \hat{\mathbf{a}}, \quad \cos\theta_- = -\hat{\boldsymbol{\ell}}_- \cdot \hat{\mathbf{b}}. \quad (7)$$

Using the definitions (7), integrating out the azimuthal angles in Eq. (6) and separating it into single-differential cross sections yields the following three relations between angular observables and parameters of $\mathbf{B}^\pm$ and $C$:

$$\frac{1}{\sigma}\frac{d\sigma}{d\cos\theta_\pm^i} = \frac{1}{2}\left(1 + b_i^\pm \cos\theta_\pm^i\right) \quad (8)$$

$$\frac{1}{\sigma}\frac{d\sigma}{d\cos\theta_+^i \cos\theta_-^i} = \frac{1}{2}\left(1 - c_{ii}\cos\theta_+^i \cos\theta_-^i\right) \times \ln\left(|\cos\theta_+^i \cos\theta_-^i|^{-1}\right) \quad (9)$$

$$\frac{1}{\sigma}\frac{d\sigma}{d\zeta} = \frac{1}{2}\left(1 - \frac{C_{ij} \pm C_{ji}}{2}\zeta\right)\arccos|\zeta|$$

with: $\zeta = \cos\theta_+^i \cos\theta_-^j \pm \cos\theta_+^j \cos\theta_-^i \ (i \neq j)$ (10)

While it is possible to fit the normalised differential distributions (8)-(10) to their functional form in order to extract the relevant polarisation and spin coefficients, one can also take the expectation value of any of these observables $\xi$ to determine $\mathbf{B}^\pm$ and $C$:

$$\langle\xi\rangle = \frac{1}{\sigma}\int d\sigma\, \xi, \quad \mathbf{B}^\pm = 3\langle\xi\rangle, \quad C = -9\langle\xi\rangle. \quad (11)$$

An additional observable can be defined that exhibits sensitivity to the diagonal elements of the spin correlation matrix:

$$\cos\varphi = \hat{\boldsymbol{\ell}}_+ \cdot \hat{\boldsymbol{\ell}}_-, \quad (12)$$

for which the normalised differential distribution obeys:

$$\frac{1}{\sigma}\frac{d\sigma}{d\cos\varphi} = \frac{1}{2}(1 - D\cos\varphi), \quad (13)$$

and $D = -\frac{1}{3}\mathrm{Tr}\, C$. This coefficient can similarly be obtained by taking the expectation value of the opening angle distribution, as $D = -3\langle\cos\varphi\rangle$.

### 2.2 Predictions at NLO QCD

In order to access the detailed kinematics of the top quarks and $Z$ bosons, as well as those of their decay products, we use the MADGRAPH 2.8.1 generator [23] in NLO+PS mode, interfaced to PYTHIA 8.244 [24]. We generate ten million $t\bar{t}Z$ events at NLO precision in QCD, using the NNPDF3.0 PDF set [25] in the five-flavour scheme and with the following parameters:

$m(t) = 172.5\,\mathrm{GeV}, \quad \Gamma_t = 1.3197\,\mathrm{GeV},$
$m(W) = 80.399\,\mathrm{GeV}, \quad \Gamma_W = 2.085\,\mathrm{GeV},$
$m(Z) = 91.1876\,\mathrm{GeV}, \quad \Gamma_Z = 2.4952\,\mathrm{GeV},$
$G_F = 1.16637 \times 10^{-5}\,\mathrm{GeV}^{-2}, \quad \alpha_s(m_Z) = 0.118,$
$\mu_R = \mu_F = m_t + m_Z.$

We use MadSpin [26,27] to decay the $t\bar{t}$ system to its semi- or di-leptonic final state, explicitly setting the parameter spinmode to the value madspin in order to retain the full spin correlations. A second, statistically independent $t\bar{t}Z$ sample is generated in the exact same way but for this parameter, now set to none. These samples are hereafter referred to as "spin-on" and "spin-off".

---
[2] We note that the early ATLAS analysis [22] attempts such a measurement of $t\bar{t}$ spin correlations in the $\ell$+jets channel, but is strongly limited by the absence of $c$-tagging, relying instead on a 2D map of $b$-tagging weights versus $p_T$.

**Table 1** Spin correlation, cross-correlation and top polarisation coefficients in the SM, extracted at LO and NLO QCD precision for $t\bar{t}Z$ events at $\sqrt{s} = 13, 14$ TeV. The central values correspond to a choice of scales $\mu_R = \mu_F = m_t + m_Z$, while the standard deviations reflect both the Monte Carlo statistical uncertainties and variations of the scales (as described in the text). The fourth and sixth columns display the results of a similar calculation in $t\bar{t}$ events at mixed NLO QCD and electroweak precision ("NLOW"); these numbers are quoted from Ref. [20]

| Coefficient | $t\bar{t}Z$ LO 13 TeV | $t\bar{t}Z$ NLO 13 TeV | $t\bar{t}$ NLOW 13 TeV | $t\bar{t}Z$ NLO 14 TeV | $t\bar{t}$ NLOW 14 TeV |
|---|---|---|---|---|---|
| $c_{rr}$ | $-0.207 \pm 0.007$ | $-0.198 \pm 0.009$ | $0.071 \pm 0.008$ | $-0.190 \pm 0.009$ | $0.072 \pm 0.008$ |
| $c_{kk}$ | $-0.197 \pm 0.013$ | $-0.193 \pm 0.009$ | $0.331 \pm 0.002$ | $-0.182 \pm 0.009$ | $0.331 \pm 0.002$ |
| $c_{nn}$ | $-0.125 \pm 0.003$ | $-0.117 \pm 0.005$ | $0.326 \pm 0.002$ | $-0.118 \pm 0.005$ | $0.325 \pm 0.002$ |
| $c_{rk}$ | $-0.163 \pm 0.003$ | $-0.173 \pm 0.007$ | $-0.206 \pm 0.002$ | $-0.180 \pm 0.007$ | $-0.204 \pm 0.004$ |
| $c_{kn}$ | $0.000 \pm 0.003$ | $0.012 \pm 0.006$ | $\lesssim 2 \times 10^{-3}$ | $-0.001 \pm 0.006$ | $\lesssim 2 \times 10^{-3}$ |
| $c_{rn}$ | $0.003 \pm 0.003$ | $-0.004 \pm 0.006$ | $\lesssim 1 \times 10^{-3}$ | $0.006 \pm 0.006$ | $\lesssim 1 \times 10^{-3}$ |
| $c_r$ | $0.008 \pm 0.003$ | $0.007 \pm 0.007$ | $\lesssim 1 \times 10^{-3}$ | $-0.004 \pm 0.007$ | $\lesssim 1 \times 10^{-3}$ |
| $c_k$ | $-0.003 \pm 0.003$ | $0.003 \pm 0.006$ | $\lesssim 1 \times 10^{-3}$ | $0.001 \pm 0.006$ | $\lesssim 1 \times 10^{-3}$ |
| $c_n$ | $0.001 \pm 0.003$ | $0.005 \pm 0.007$ | $\lesssim 1 \times 10^{-3}$ | $-0.008 \pm 0.007$ | $\lesssim 1 \times 10^{-3}$ |
| $b_r^+$ | $0.058 \pm 0.003$ | $0.055 \pm 0.002$ | $\lesssim 2 \times 10^{-3}$ | $0.055 \pm 0.002$ | $\lesssim 2 \times 10^{-3}$ |
| $b_r^-$ | $0.060 \pm 0.003$ | $0.055 \pm 0.002$ | $\lesssim 2 \times 10^{-3}$ | $0.057 \pm 0.002$ | $\lesssim 2 \times 10^{-3}$ |
| $b_k^+$ | $-0.069 \pm 0.002$ | $-0.077 \pm 0.002$ | $\lesssim 4 \times 10^{-3}$ | $-0.077 \pm 0.002$ | $\lesssim 4 \times 10^{-3}$ |
| $b_k^-$ | $-0.071 \pm 0.002$ | $-0.076 \pm 0.002$ | $\lesssim 4 \times 10^{-3}$ | $-0.074 \pm 0.002$ | $\lesssim 4 \times 10^{-3}$ |
| $b_n^+$ | $-0.001 \pm 0.001$ | $0.001 \pm 0.002$ | $\lesssim 3 \times 10^{-3}$ | $0.001 \pm 0.002$ | $\lesssim 3 \times 10^{-3}$ |
| $b_n^-$ | $0.001 \pm 0.001$ | $0.001 \pm 0.002$ | $\lesssim 3 \times 10^{-3}$ | $-0.001 \pm 0.002$ | $\lesssim 3 \times 10^{-3}$ |

We construct the normalised differential distributions from Eqs. (8)–(10). From the properly weighted expectation values of these distributions, the three diagonal spin correlation, six cross-correlation and six top polarisation coefficients are determined and reported in Table 1, for a center-of-mass energy $\sqrt{s} = 13$ TeV. A single uncertainty is quoted on each extracted component of the spin density matrix, corresponding to the sum in quadrature of the Monte Carlo statistical uncertainty as well as up and down variations of the renormalisation and factorisation scales ($\mu_R$ and $\mu_F$). Following the usual convention of the ATLAS and CMS collaborations, we consider $\mu_R = \{0.5, 2\} \times \mu_0$, $\mu_F = \{0.5, 2\} \times \mu_0$ and $\mu_R = \mu_F = \{0.5, 2\} \times \mu_0$ as three independent sources of theoretical uncertainty, where $\mu_0 = m_t + m_Z$. For most coefficients, except $c_{rr}$ and $c_{kk}$, the total scale uncertainty is largely subdominant with respect to the statistical uncertainty on the generated Monte Carlo sample. Uncertainties arising from the PDF are known to have a negligible impact [14,15], compared to the above scale uncertainties, and are not considered here. In order to ease the comparison between our predictions for the $t\bar{t}Z$ spin density matrix and the previous results for the $t\bar{t}$ case, we also include in Table 1 the mixed NLO QCD and electroweak ("NLOW") calculations from Tables 7 and 8 of Ref. [20].

The three diagonal spin correlation coefficients ($c_{rr}$, $c_{kk}$ and $c_{nn}$) are all non-zero in both $t\bar{t}$ and $t\bar{t}Z$ events, but adopt radically different values. While at tree-level the $c_{kk}$ coefficient is mostly driven by the $q\bar{q} \to t\bar{t}Z$ diagrams (where the $Z$ boson can be radiated in the initial state), both $c_{rr}$ and $c_{nn}$ are found to be more sensitive to the gluon-initiated channel (where the $Z$ boson is necessarily emitted from one of the top quarks). All three coefficients receive small positive corrections at NLO, from the opening of the mixed $qg$ initial state. The contribution of $q\bar{q}$- and $qg$-initiated diagrams to the $t\bar{t}Z$ process is slightly higher than for $t\bar{t}$, since the $Z$ boson couples more strongly to down-type quarks.

As is the case for $t\bar{t}$ and as was argued at the beginning of Sect. 2, all cross-correlation coefficients are suppressed in the SM, except for $c_{rk}$. The transverse top polarisations $b_n^\pm$ are similarly compatible with zero and in agreement with the $t\bar{t}$ predictions. However, the emission of the $Z$ boson from either the top or anti-top quark induces a small longitudinal polarisation which is reflected in non-zero $b_r^\pm$ and $b_k^\pm$ coefficients. While this effect is much smaller than the size of the spin correlations, it could, once future $t\bar{t}Z$ measurements have reached the necessary precision, be used as a further handle on hypothetical CP-violating $t$-$Z$ couplings.

In Table 1, we also report the results of the same calculation at a higher center-of-mass energy of $\sqrt{s} = 14$ TeV. As was the case for $t\bar{t}$ in Ref. [20], we find figures consistent with those quoted at $\sqrt{s} = 13$ TeV, within uncertainties. Using the alternative "spin-off" sample, we verified that all the coefficients of the resulting $t\bar{t}Z$ spin density matrix were compatible with zero, as expected. Results are further presented at LO QCD precision for $\sqrt{s} = 13$ TeV only, and found to be close to the NLO QCD values.

## 3 Towards an observation of spin correlation in $t\bar{t}Z$ events at the LHC

We now turn to the potential of the ATLAS and CMS experiments to measure the coefficients of the $t\bar{t}Z$ spin density matrix in LHC data. In Sect. 3.1, we briefly summarise the current experimental precision and consider the possibility of a proxy observable sensitive to an overall "fraction of SM-like spin correlation", $f_{SM}$. In Sect. 3.2, we show that, according to our evaluation, current LHC Run 2 $t\bar{t}Z$ measurements would only be able to show evidence (at the $3\sigma$ level) of spin correlations if the full spin density matrix is measured together with the highly sensitive $\cos\varphi$ observable. We then make projections for future runs of the LHC, and find that an observation ($5\sigma$) can be made with the combined Run 2 and Run 3 datasets.

### 3.1 Observables and experimental uncertainties

Recent measurements of the inclusive and differential $t\bar{t}Z$ cross section by the ATLAS and CMS collaborations [14,15] are performed exclusively in the trilepton ($3\ell$) and tetralepton ($4\ell$) channels, where the $t\bar{t}$ system decays semi- and di-leptonically, respectively. The di-leptonically decaying $t\bar{t}$ in the tetralepton final state may be used for measurements of spin correlation and polarisaton, whereas the semi-leptonically decaying $t\bar{t}$ can only be used for polarisation measurements. These multi-lepton selections ensure a high purity of signal events as well as an accurate reconstruction of key kinematic observables due to the excellent reconstruction of electrons and muons in both the ATLAS and CMS detectors.

The ATLAS analysis [14] uses the full LHC Run 2 dataset, corresponding to an integrated luminosity of 139.1 fb$^{-1}$, and reports about 430 $t\bar{t}Z$ events in their $3\ell$ selection and about 90 events in their $4\ell$ selection. The CMS analysis [15], on the other hand, only uses the dataset collected during the 2016-2017 period and amounting to an integrated luminosity of 77.5 fb$^{-1}$. Their higher signal acceptance provides them with almost 400 $t\bar{t}Z$ events in the $3\ell$ channel and about 60 events in the $4\ell$ one. It is therefore reasonable to assume that a legacy Run 2 analysis, benefiting from more constrained backgrounds and higher acceptances, could provide 500 signal events (with reconstructed top quarks) in the trilepton final state, and 100 signal events in the tetralepton final state. We therefore derive our expected statistical uncertainties based on the assumption that 100 events can be used to determine the spin correlation and cross-correlation coefficients, and 600 events for the individual top polarisation coefficients.[3] From the ATLAS measurement [14], and in particular the normalised differential cross section distribution for $|\Delta\phi(\ell_t^+, \ell_{\bar{t}}^-)|$ and $p_T(t\bar{t})$ in the $4\ell$ parton-level fiducial volume (Figures 14b, 14d, 15b and 15d of Ref. [14]), we estimate that an 8% total systematic uncertainty can be assigned to each of 4 bins (equally dividing the range [−1, 1]) for all the angular distributions we consider in this study.

The $|\Delta\phi(\ell_t^+, \ell_{\bar{t}}^-)|$ observable, the opening angle between the two charged leptons from the decay of the $t\bar{t}$ system in the laboratory frame, was considered as a proxy observable in the ATLAS [5] and CMS [4] measurements of spin correlation in dileptonic $t\bar{t}$ events. A clear difference in shape was exhibited between Monte Carlo templates with and without SM-like spin correlations. This observable does not require the full reconstruction of the $t\bar{t}$ system, which in the dilepton final state is necessarily an underconstrained problem (since one must solve for the two neutrino momenta using the single quantity that is the missing transverse energy in the event). In Fig. 1, we show the corresponding distribution constructed from $t\bar{t}Z$ events in the $4\ell$ final state. The bottom panel of Fig. 1, displaying the ratio of the spin-on to spin-off hypotheses, clearly does not exhibit the strong monotonic slope observed in Ref. [5], and therefore $|\Delta\phi(\ell_t^+, \ell_{\bar{t}}^-)|$ can no longer be considered a viable proxy observable. This is due to the additional radiation of the $Z$ boson from either the initial or the final state, disrupting the balance of the nominal $t\bar{t}$ event and washing out the spin information in the laboratory frame. Full reconstruction of the $t\bar{t}$ system is therefore required in order to access the elements of the spin density matrix, and the $\cos\varphi$ observable (defined by Eq. (12) and relying on boosts of the leptons to the rest frames of their parent top quarks) becomes accessible. Fig. 2 shows its corresponding distribution in $4\ell$ $t\bar{t}Z$ events, where the linear slope induced by the presence of non-zero spin correlations and predicted by Eq. (13) makes it an ideal candidate for a template fit.

### 3.2 Template fit and extrapolations

Using the two Monte Carlo samples described in Sect. 2.2, we can produce for any observable $\mathcal{O}$ two predictions at NLO accuracy in QCD, one taking into account SM-like spin correlations ($\mathcal{O}_{\text{spin-on}}$) and one neglecting them completely ($\mathcal{O}_{\text{spin-off}}$). We now turn to the extraction of a single parameter of interest $f_{SM}$ from template fits to various choices of $\{\mathcal{O}\}$, writing:

---

[3] The dependence of the results presented in Fig. 4 on this assumption of events statistics was tested: we found that the results were only significantly degraded if our assumption was an overestimation by 33% of the real number of events achievable

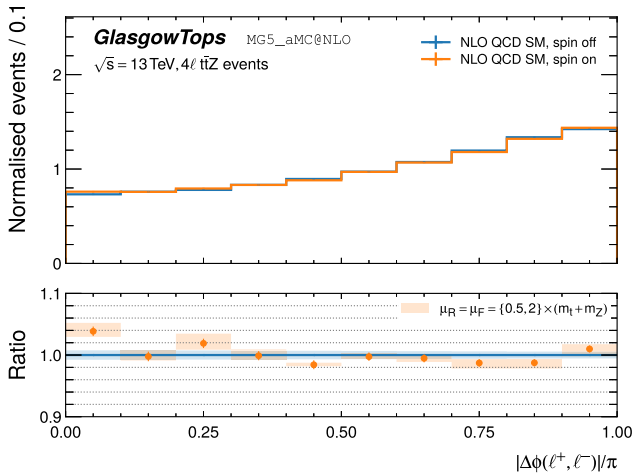

**Fig. 1** Normalised distribution of $|\Delta\phi(\ell_t^+, \ell_{\bar{t}}^-)|$ in $4\ell$ $t\bar{t}Z$ events, under the assumption of SM-like (orange) and no spin correlations (blue). In the bottom panel, the orange error bars and the blue error band represent the Monte Carlo statistical uncertainties on these two distributions. The orange error band further corresponds to the scale uncertainty obtained by varying $\mu_R$ and $\mu_F$ together by a factor of 2

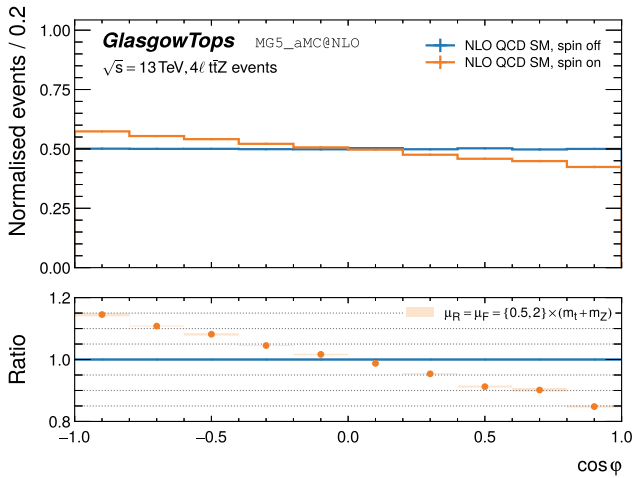

**Fig. 2** Normalised distribution of $\cos\varphi$ in $4\ell$ $t\bar{t}Z$ events, under the assumption of SM-like (orange) and no spin correlations (blue). In the bottom panel, the orange error bars and the blue error band represent the Monte Carlo statistical uncertainties on these two distributions. The orange error band further corresponds to the scale uncertainty obtained by varying $\mu_R$ and $\mu_F$ together by a factor of 2

$$\mathcal{O} = f_{\text{SM}} \cdot \mathcal{O}_{\text{spin-on}} + (1 - f_{\text{SM}}) \cdot \mathcal{O}_{\text{spin-off}}. \quad (14)$$

We use EFT*fitter* [28], based on the *Bayesian Analysis Toolkit* BAT.jl package [29], to perform the template fit in the context of Bayesian statistics. A flat prior is set for $f_{\text{SM}}$, and we construct a pseudo-dataset in the $4\ell$ final state, according to the $\mathcal{O}_{\text{spin-on}}$ template. Considering four bins of equal width in the range $[-1, 1]$ for the observable $\cos\varphi$, we assume the statistical and systematic uncertainties (as defined in the previous section) on this pseudo-measurement to follow a Gaussian distribution; statistical correlations between bins are taken into account in the model. Based on our previously discussed assumptions for a suitable Run 2 measurement, we determine a 90% credibility interval for $f_{\text{SM}}$ to be [0.12, 1.88]. The associated Bayes factor, quantifying the rejection of the null (no-spin) hypothesis, is 1.44; according to the qualitative classification of Ref. [30], this is only "anecdotal evidence" in favour of non-zero $t\bar{t}Z$ spin correlations. If one repeats the template fit, but this time considering only the extracted coefficients of the full spin density matrix, a tighter interval of [0.24, 1.76] is found – with a corresponding Bayes factor of 3.83, this is now "substantial evidence" against the null hypothesis. Only by combining these two sources of information (the normalised differential distribution of $\cos\phi$ and the $3 + 6 + 6$ spin correlation and top polarisation coefficients) do we achieve a 90% credibility interval of [0.46, 1.54], and a Bayes factor of 13.5 or "strong evidence" (comparable to the $3\sigma$ level of statistical significance in the frequentist approach commonly employed in both the ATLAS and CSM collaborations).

Having defined these three possible template fits to $f_{\text{SM}}$, we explore further scenarios. Assuming datasets with integrated luminosities of 300 fb$^{-1}$ and 3000 fb$^{-1}$ are to be collected during Run 3 and the entirety of HL-LHC operations respectively, we scale the expected $t\bar{t}Z$ event yields appropriately. Based on the calculations of Ref. [16], the inclusive $t\bar{t}Z$ cross section increases with the center-of-mass energy, by a factor of $\sim 1.21$ when going from $\sqrt{s} = 13$ TeV to $\sqrt{s} = 14$ TeV. We assume that data taking during both Run 3 and HL-LHC will be at $\sqrt{s} = 14$ TeV, and consider this cross section enhancement in our estimations. We further consider a scenario where the Run 2 and Run 3 measurements are combined, and one where the HL-LHC measurement eventually benefits from systematic uncertainties reduced by a factor of two. The resulting 90% Bayesian credibility intervals on $f_{\text{SM}}$ are presented in Table 2 and Fig. 3, while the corresponding Bayes factors are shown in Fig. 4. At the level of the combination of the Run 2 and Run 3 measurements of both the full spin density matrix and the $\cos\varphi$ observable, a meaningful claim for an observation of spin correlations in $t\bar{t}Z$ events can be made (the $5\sigma$ "golden standard" in high energy physics frequentism). Should the legacy ATLAS and CMS Run 2 $t\bar{t}Z$ analyses exceed our expectations, their combination would achieve this milestone before the end of Run 3.

## 4 SMEFT interpretation

The popular framework of the SMEFT extends the SM Lagrangian as [31]

$$\mathcal{L}_{\text{SMEFT}} = \mathcal{L}_{\text{SM}} + \sum_{d>4} \mathcal{L}^{(d)}, \quad \mathcal{L}^{(d)} = \sum_{i=1}^{n_d} \frac{C_i^{(d)}}{\Lambda^{d-4}} \mathcal{Q}_i^{(d)}, \quad (15)$$

**Table 2** 90% credibility intervals on $f_{\mathrm{SM}}$, according to the various scenarios described in the text of Sect. 3.2. Here, "HL-LHC ⊕" refers to the HL-LHC setup with improved systematic uncertainties

| Scenario | $\cos\varphi$ only | $C/\mathbf{B}^{\pm}$ only | full information |
| --- | --- | --- | --- |
| Run 2 | [0.12, 1.88] | [0.24, 1.76] | [0.46, 1.54] |
| Run 3 | [0.17, 1.84] | [0.37, 1.62] | [0.58, 1.43] |
| Run 2 + Run 3 | [0.20, 1.82] | [0.49, 1.52] | [0.66, 1.33] |
| HL-LHC | [0.42, 1.58] | [0.58, 1.43] | [0.68, 1.31] |
| HL-LHC ⊕ | [0.58, 1.43] | [0.74, 1.26] | [0.82, 1.18] |

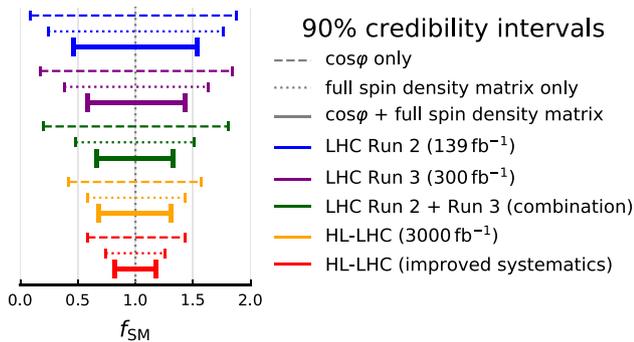

**Fig. 3** Comparison of the 90% credibility intervals on $f_{\mathrm{SM}}$, according to the various scenarios described in the text of Sect. 3.2

building $d$-dimensional operators $\mathcal{Q}_i^{(d)}$ from the usual SM fields, to which a Wilson coefficient $C_i^{(d)}$ is associated as well as suitable power of the cutoff scale $\Lambda$. The number of such operators, $n_d$, is known up to $d = 8$; the development of appropriate numerical tools in recent years now allows for a detailed study of SMEFT phenomenology at the LHC up to $d = 6$ in the Warsaw basis. In this work, we consider only contributions from $\mathcal{L}^{(6)}$, since $\mathcal{L}^{(5)}$ is well known to include baryon- and lepton-number violating operators, while higher order operators are $\Lambda$-suppressed. We further adopt the conventional choice of scale $\Lambda = 1$ TeV, such that our numerical results can straightforwardly be recast to a different scale or interpreted as limits on the dimensionful coefficients $\tilde{C}_i \equiv C_i/\Lambda^2$ in units of TeV$^{-2}$.

In Sect. 4.1, we define the SMEFT operators relevant to the $t\bar{t}Z$ interaction vertex and their relation to the anomalous coupling formalism. The setup of our calculation is detailed in Sect. 4.2, and the results of fits to the angular observables of Sect. 2.1 are presented in 4.3. Disregarding information about the total production rates entirely, we find new sensitivity to these SMEFT operators in some of the spin correlation observables.

4.1 Formalism of the top-$Z$ interaction

The coupling of the top quark to the electroweak bosons and the gluon can be described with the following operators [32]:

$$\begin{aligned}
\mathcal{Q}_{uG}^{(ij)} &= \left(\bar{q}_i \sigma^{\mu\nu} T^A u_j\right) \tilde{\varphi} G_{\mu\nu}^A, \\
\mathcal{Q}_{uB}^{(ij)} &= \left(\bar{q}_i \sigma^{\mu\nu} u_j\right) \tilde{\varphi} B_{\mu\nu}, \\
\mathcal{Q}_{uW}^{(ij)} &= \left(\bar{q}_i \sigma^{\mu\nu} \tau^I u_j\right) \tilde{\varphi} W_{\mu\nu}^I, \\
\mathcal{Q}_{\varphi u}^{(ij)} &= \left(\varphi^\dagger i\overset{\leftrightarrow}{D}_\mu \varphi\right) \left(\bar{u}_i \gamma^\mu u_j\right), \\
\mathcal{Q}_{\varphi q}^{1(ij)} &= \left(\varphi^\dagger i\overset{\leftrightarrow}{D}_\mu \varphi\right) \left(\bar{q}_i \gamma^\mu q_j\right), \\
\mathcal{Q}_{\varphi q}^{3(ij)} &= \left(\varphi^\dagger i\overset{\leftrightarrow}{D}_\mu^I \varphi\right) \left(\bar{q}_i \gamma^\mu \tau^I q_j\right),
\end{aligned} \qquad (16)$$

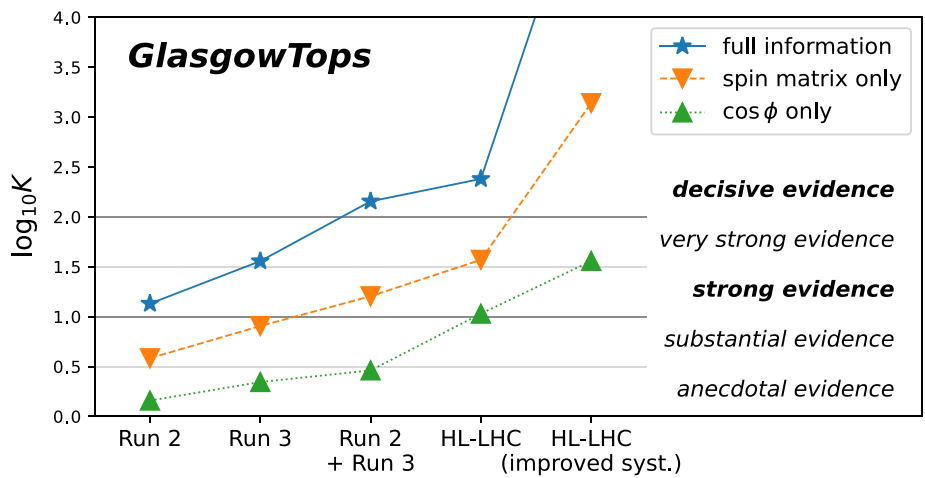

**Fig. 4** Evolution of the Bayes factor $K$ characterising the rejection of the no-spin hypothesis for the various scenarios described in the text of Sect. 3.2. The qualitative description for various ranges of $K$ is taken from Ref. [30]; the ones in bold correspond to the $3\sigma$ and $5\sigma$ frequentist standards. The rightmost blue marker is omitted from the Figure, as we find $\log_{10} K \gg 2$, indicating that this particular scenario corresponds to a precision measurement of $f_{\mathrm{SM}}$ and clearly rejects the null hypothesis (absence of $t\bar{t}Z$ spin correlations)

employing the notation of Ref. [33] where flavour indices are labelled by $ij$; left-handed quark doublets are denoted $q$ and right-handed singlets $u, d$; $\tau^I$ and $T^A \equiv \lambda^A/2$ are the Pauli and Gell-Mann matrices respectively; the Higgs doublet is $\varphi$ and $\tilde{\varphi} \equiv i\tau^2\varphi$.

We then consider the degrees of freedom listed below, and built from the Wilson coefficients corresponding to the dimension-6 operators of (16):

$$\begin{aligned}
c_{tG} &\equiv \text{Re}\left\{C_{uG}^{(33)}\right\}, \\
c_{tZ} &\equiv \text{Re}\left\{-\sin\theta_W C_{uB}^{(33)} + \cos\theta_W C_{uW}^{(33)}\right\}, \\
c_{\varphi Q}^3 &\equiv C_{\varphi q}^{3(33)}, \\
c_{\varphi Q}^- &\equiv C_{\varphi q}^{1(33)} - C_{\varphi q}^{3(33)}, \\
c_{\varphi t} &\equiv C_{\varphi u}^{(33)},
\end{aligned} \quad (17)$$

where $\theta_W$ is the Weinberg angle. We do not consider the imaginary counterparts to $c_{tG}$ and $c_{tZ}$ here, in order to restrain our interpretation to CP-even EFT effects. Furthermore, we do not include the four-quark operators that would otherwise enter in $t\bar{t}Z$ production, as these can better be constrained in $t\bar{t}$ measurements, and we neglect contributions from off-shell $t\bar{t} + \ell\ell$ production, including operators of the form $(\bar{\ell}\Gamma\ell)(\bar{t}\Gamma t)$.[4] We choose to focus in this paper on the operators listed in (17), which have previously formed the basis of the ATLAS [35] and CMS [15,36,37] $t\bar{t}Z$ EFT interpretations, and leave to future work the exhaustive study of all SMEFT contributions to $t\bar{t}Z$ spin correlations, including one-loop EFT and CP-odd effects.

Writing the generic Lagrangian for the $t\bar{t}Z$ vertex [38–40],

$$\mathcal{L}_{t\bar{t}Z} = e\bar{u}(p_t)\left[\gamma^\mu(C_{1,V} + \gamma_5 C_{1,A}) + \frac{i\sigma^{\mu\nu}q_\nu}{m_Z}(C_{2,V} + i\gamma_5 C_{2,A})\right]v(p_{\bar{t}})Z_\mu, \quad (18)$$

with $q_\nu \equiv (p_t - p_{\bar{t}})_\nu$, the set of Wilson coefficients above can be translated into anomalous couplings [41]

$$\begin{aligned}
C_{1,V} &= C_V^{\text{SM}} + \frac{v^2}{2\Lambda^2 \sin\theta_W \cos\theta_W}\text{Re}\left\{-c_{\varphi t} - c_{\varphi Q}^-\right\}, \\
C_{1,A} &= C_A^{\text{SM}} + \frac{v^2}{2\Lambda^2 \sin\theta_W \cos\theta_W}\text{Re}\left\{-c_{\varphi t} + c_{\varphi Q}^-\right\}, \\
C_{2,V} &= \frac{\sqrt{2}v^2}{2\Lambda^2 \sin\theta_W \cos\theta_W}c_{tZ}, \\
C_{2,A} &= \frac{\sqrt{2}v^2}{2\Lambda^2 \sin\theta_W \cos\theta_W}c_{tZ}^I,
\end{aligned} \quad (19)$$

where $c_{tZ}^I$ is the imaginary counterpart to $c_{tZ}$ and $v \approx 246$ GeV is the Higgs vev. The weak magnetic and electric dipole moments of the top quark, $C_{2,V}$ and $C_{2,A}$ respectively, are both highly suppressed in the SM [38,40]. The latter is CP-violating and can therefore be sourced by $c_{tZ}^I$, though we choose to ignore such effects here. The CMS collaboration in particular reports limits on anomalous couplings [15,36].

### 4.2 Setup of the reinterpretation

We generate a Monte Carlo sample of 10 million $t\bar{t}Z$ events at parton-level with MADGRAPH and PYTHIA 8, using the same parameters as in Sect. 2.2 but including the dim6top_LO UFO model [42] and reducing the precision to LO QCD. We then use the reweighting module of MADGRAPH [43] to probe multiple values of the five Wilson coefficients listed in Eq. 17. Only single insertions are allowed at the amplitude level in $t\bar{t}Z$ production. The dependence of a generic observable $\mathcal{O}$ (e.g. an inclusive or differential cross section, or in our case an expectation value) on a set of EFT operators associated to Wilson coefficients $\{\tilde{C}_i\}$ can be parameterised quadratically as:

$$\mathcal{O} = \mathcal{O}_{\text{SM}} + \sum_i \tilde{C}_i A_i + \sum_{i,j} \tilde{C}_i \tilde{C}_j B_{ij}, \quad (20)$$

such that our reweighting approach allows us to readily extract the SM-EFT interference terms $A_i$ and pure EFT terms $B_{ij}$. When building the normalised differential distributions corresponding to the angular observables, we divide the SMEFT prediction in each bin, $\frac{d\sigma}{dX^i}$, by the integral over all bins, $\sum_i \frac{d\sigma}{dX^i}$. In this way, only the shape effects induced by the EFT component come into play, and not the absolute cross section. Since the latent SM prediction for each observable, $\mathcal{O}_{\text{SM}}$, is only at LO precision in QCD, we adopt the convention of the LHC Top Working Group and apply an overall rescaling $\mathcal{O} \to \mathcal{O} \times \mathcal{O}_{\text{SM}}^{\text{best}}/\mathcal{O}_{\text{SM}}$ to the best available SM prediction (for the purposes of this work, this is the NLO QCD calculation of Sect. 2.2). In the case of the $\cos\varphi$ distribution, this rescaling is applied bin by bin.

No significant sensitivity (with respect to the numerical precision achieved) is found for $c_{\varphi Q}^3$ in any of the angular distributions we have considered so far; the dependence on $c_{\varphi Q}^-$ is also very weak. Both these observations can be understood in light of the very tight constraints on a particular combination[5] of these two operators already imposed by LEP measurements of the $bbZ$ coupling [44]. Due to its very small effect on the selected observables, completely covered by the associated Monte Carlo statistical uncertainty, we eliminate $c_{\varphi Q}^3$ from our analysis.

---

[4] In the phase-space $|m_{\ell\ell} - m_Z| \leq 10$ GeV commonly defined in both ATLAS and CMS analyses, these operators do not interfere with the SM and only enter at $\mathcal{O}(\Lambda^{-4})$ [34].

[5] Namely $c_{\varphi Q}^- + 2c_{\varphi Q}^3$, often denoted by $c_{\varphi Q}^+$ in the literature.

**Table 3** Exclusion bounds on the four operators of interest reported by the SMEFiT global analysis as well as previous $t\bar{t}Z$ measurements by the ATLAS and CMS collaborations, as 95% CL frequentist intervals. Limits from our Bayesian fit, for the two scenarios described in the accompanying text, are given as 95% credibility intervals, assuming informative priors on all the Wilson coefficients

| 95% bounds | | $c_{tZ}$ | $c_{\varphi Q}^-$ | $c_{tG}$ | $c_{\varphi t}$ |
|---|---|---|---|---|---|
| SMEFiT | [45] | [−4.6, 5.9] | [−2.6, 3.3] | [−0.4, 0.4] | [−23, 7.3] |
| ATLAS 36 fb$^{-1}$ | [35] | [−4.9, 4.9] | [−3.3, 4.2] | – | [−25, 5.5] |
| CMS 45 fb$^{-1}$ | [37] | [−3.3, 3.2] | [−7.6, 22] | [−1.4, 1.2] | [−19, 12] |
| CMS 77.5 fb$^{-1}$ | [15] | [−1.1, 1.1] | [−4.0, 0.0] | – | [0.3, 5.4] |
| Run 2 | | [−2.7, 3.4] | [−2.6, 2.6] | [−0.4, 0.4] | [−6.0, 5.6] |
| Run 2 + Run 3 | | [−2.2, 2.8] | [−2.6, 2.6] | [−0.4, 0.4] | [−5.0, 4.6] |

Our SMEFT interpretation is based on the EFT*fitter* framework [28], and we use the same projected uncertainties and correlations as previously described in Sect. 3. We consider all the top polarisation and spin correlation coefficients, as well as four bins of the $\cos\varphi$ distribution, and restrict ourselves to the Run 2 and Run 2 + Run 3 combination scenarios. Priors are set for each Wilson coefficient to reflect current limits from global analyses; we elect to use Gaussian distributions with mean zero and width half the LO QCD marginalised bounds reported by the SMEFiT collaboration at the 95% CL (Table 5.3 of Ref. [45]). Note that while we do not rely on the inclusive cross section (total rate) information in this work, the exclusion limits set in the SMEFiT paper do. We do not consider uncertainties on the EFT paremeterisation, either from scale uncertainties or missing dimension-8 terms, as these are usually much smaller than the statistical and systematic uncertainties on the experimental measurements.

### 4.3 New constraints from angular observables

We perform a global fit to the four Wilson coefficients under consideration and present the marginalised 95% credibility intervals in Table 3. For reference, we also include the 95% CL bounds from the SMEFiT analysis [45], as well as from several $t\bar{t}Z$ measurements by the ATLAS [35] and CMS collaborations [15,37]. We observe that the credibility intervals we derive for $c_{\varphi Q}^-$ and $c_{tG}$ in both the Run 2 and Run 2 + Run 3 combination scenarios correspond to the informed priors we used; in other words, the expected precision of the underlying measurements is too low to provide additional constraining power on these two operators. On the other hand, we notice an improvement in the constraining of $c_{\varphi t}$ and $c_{tZ}$, the latter appearing to be the most sensitive. Moreover, these have the potential to complement (and be somewhat competitive with) the ATLAS and CMS results that rely on the total rate information.

We repeat the fit as many times as there are observables, omitting one of those observable each time. This allows us to gauge the relative importance of each pseudo-measurement, by monitoring the increase in the exclusion bounds. This approach reveals that, as expected, the diagonal spin correlation coefficients ($c_{rr}$, $c_{kk}$ and $c_{nn}$) are amongst the most sensitive observables, together with the two extreme bins of the $\cos\varphi$ differential distribution. Of the top polarisation coefficients, $b_k^\pm$ are the most important ones to include in the fit, whilst $b_n^\pm$ has an almost negligible contribution to the constraining power.

We perform a further iteration of the fit, using uniform priors (in the range [−10, 10]) for all the Wilson coefficients. The corresponding results, reported in Table 4, show slightly degraded bounds on $c_{tZ}$ and $c_{\varphi t}$ when compared to the original fit. This is due to the $c_{\varphi Q}^-$ and $c_{tG}$ coefficients now being properly constrained. As previously claimed, there is little effective dependence of any of the observables we consider on $c_{\varphi Q}^-$, but there is a much more remarkable effect on $c_{tG}$. While still far from the results of the SMEFiT analysis [45] (which mostly leverages $t\bar{t}$ production rates), our bounds on $c_{tG}$ are closer to the limits reported by the CMS analyses – but rely only on normalised, angular shape effects.

Following the work of Ref. [46], we compute the covariance matrix for the fit with uniform priors, and take its inverse to be an approximation of the Fisher information matrix. Large entries of the Fisher matrix therefore correspond to well-measured directions in the space spanned by the four Wilson coefficients. Furthermore, its eigenvectors reflect the actual linear combinations of SMEFT operators being probed, while their corresponding eigenvalues determine the expected sensitivity along these directions. Denoting the former by $\mathcal{F}_i$ and the latter by $\lambda_i$, we find:

$$\begin{aligned}
\lambda_1 &= 0.03, \ \mathcal{F}_1 : 0.84 \cdot c_{\varphi Q}^- - 0.54 \cdot c_{\varphi t}, \\
\lambda_2 &= 0.12, \ \mathcal{F}_2 : 0.48 \cdot c_{\varphi Q}^- + 0.79 \cdot c_{\varphi t} + 0.37 \cdot c_{tZ}, \\
\lambda_3 &= 0.69, \ \mathcal{F}_3 : 0.21 \cdot c_{\varphi Q}^- + 0.26 \cdot c_{\varphi t} - 0.86 \cdot c_{tZ} - 0.39 \cdot c_{tG}, \\
\lambda_4 &= 0.83, \ \mathcal{F}_4 : 0.11 \cdot c_{\varphi Q}^- + 0.14 \cdot c_{\varphi t} - 0.34 \cdot c_{tZ} + 0.92 \cdot c_{tG}.
\end{aligned} \tag{21}$$

This Fisher information analysis confirms what we have observed: $\mathcal{F}_4$, having the largest eigenvalue, is the most sensitive direction, and is dominated by $c_{tG}$; hence the results of Table 4. Likewise, $c_{tZ}$ is well represented along both Fisher-rotated directions $\mathcal{F}_3$ and $\mathcal{F}_4$, which have the largest eigenvalues, and is therefore the second-best constrained coefficient. On the other hand, dependence to $c_{\varphi Q}^{-}$ is found mostly in $\mathcal{F}_1$ and the corresponding $1/\sqrt{\lambda_1}$ leads to very weak exclusion bounds. The eigenvectors of (21) clearly show a non-negligible amount of degeneracy in the SMEFT dependence of the observables we have considered: by making these directions explicit, we provide the necessary information in adding another set of measurements to this one (or designing new observables), in order to break these degeneracies.

**Table 4** Exclusion bounds on the four operators of interest, given as 95% Bayesian credibility intervals. The two scenarios quoted are described in the accompanying text, and uniform priors are chosen for all the Wilson coefficients

| 95% bounds    | $c_{tZ}$      | $c_{\varphi Q}^{-}$ | $c_{tG}$      | $c_{\varphi t}$ |
|---------------|---------------|---------------------|---------------|-----------------|
| Run 2         | [−3.1, 4.4]   | [−10, 8.1]          | [−2.0, 3.4]   | [−9.8, 7.0]     |
| Run 2 + Run 3 | [−2.5, 3.5]   | [−10, 7.0]          | [−1.4, 2.8]   | [−8.1, 6.4]     |

## 5 Conclusions

In this paper, we have for the first time studied the phenomenology of spin correlations in the $t\bar{t}Z$ process at the LHC. In Sect. 2, we have produced a complete set of predictions for the $t\bar{t}Z$ spin density matrix at NLO precision in QCD, for center-of-mass energies of 13 and 14 TeV. We have found these results to be significantly different from those previously obtained for the $t\bar{t}$ process; as such, a measurement by the ATLAS and CMS experiments of spin and polarisation observables in $t\bar{t}Z$ events would be a further test of the particularly important connections between the top quark and the electroweak sectors of the Standard Model, in one of the heaviest processes that can be produced at the LHC.

Indeed in Sect. 3, we have offered a simple analysis strategy that builds directly on the existing experimental efforts of the ATLAS and CMS collaborations. By performing a template fit to the coefficients of the $t\bar{t}Z$ spin density matrix, readily extracted from a set of distinct angular observables, it is estimated that the existing LHC Run 2 dataset could be enough to claim evidence of spin correlation in $t\bar{t}Z$ events; combining these results with our predictions for Run 3, an observation could be conclusively made. Should the legacy ATLAS and CMS Run 2 analyses surpass our expectations, it is certainly possible that this milestone be achieved sooner, through their combination.

Finally, in Sect. 4 we have considered the impact of the few dimension-6 SMEFT operators relevant to $t\bar{t}Z$ production. Using only the information drawn from the spin density matrix, and neglecting any information about the total production rates, we find a novel sensitivity to the operators associated to the Wilson coefficients $c_{tZ}$ and $c_{\varphi t}$, affecting the $t$-$Z$ coupling, as well as to the top quark chromomagnetic dipole operator, which modifies the $gt\bar{t}$ vertex. We have further identified the directions in the space of SMEFT operators that are probed by our fit, using the Fisher information formalism recently applied to high energy physics.

We hope that this work can serve as the basis for an exciting new measurement of $t\bar{t} + X$ properties, which will play an important role in global EFT interpretations of the data collected at the LHC.

**Acknowledgements** BR wishes to thank Cornelius Grunwald for his technical guidance in setting up the EFT*fitter* analysis. This work was supported by the Royal Society fellowship grant URF\R1\191524.

**Data Availability Statement** This manuscript has no associated data or the data will not be deposited. [Authors' comment: No data is provided as this is phenomenological work.]

## References

1. ATLAS Collaboration, Phys. Rev. Lett. **108**, 212001 (2012). https://doi.org/10.1103/PhysRevLett.108.212001
2. ATLAS Collaboration, JHEP **03**, 113 (2017). https://doi.org/10.1007/JHEP03(2017)113
3. CMS Collaboration, Phys. Lett. B **758**, 321 (2016). https://doi.org/10.1016/j.physletb.2016.05.005
4. CMS Collaboration, Phys. Rev. D **100**(7), 072002 (2019). https://doi.org/10.1103/PhysRevD.100.072002
5. ATLAS Collaboration, Eur. Phys. J. C **80**(8), 754 (2020). https://doi.org/10.1140/epjc/s10052-020-8181-6
6. ATLAS Collaboration, Phys. Rev. Lett. **111**(23), 232002 (2013). https://doi.org/10.1103/PhysRevLett.111.232002
7. ATLAS Collaboration, Phys. Rev. D **93**(1), 012002 (2016). https://doi.org/10.1103/PhysRevD.93.012002

8. C.M.S. Collaboration, Phys. Rev. D **93**(5), 052007 (2016). https://doi.org/10.1103/PhysRevD.93.052007
9. W. Bernreuther, Z.G. Si, Nucl. Phys. B **837**, 90 (2010). https://doi.org/10.1016/j.nuclphysb.2010.05.001
10. R. Frederix, I. Tsinikos, T. Vitos, Probing the spin correlations of $t\bar{t}$ production at NLO QCD+EW (2021)
11. A. Behring, M. Czakon, A. Mitov, A.S. Papanastasiou, R. Poncelet, Phys. Rev. Lett. **123**(8), 082001 (2019). https://doi.org/10.1103/PhysRevLett.123.082001
12. M. Czakon, A. Mitov, R. Poncelet, JHEP **05**, 212 (2021). https://doi.org/10.1007/JHEP05(2021)212
13. M. Baumgart, B. Tweedie, JHEP **03**, 117 (2013). https://doi.org/10.1007/JHEP03(2013)117
14. ATLAS Collaboration, Measurements of the inclusive and differential production cross sections of a top-quark-antiquark pair in association with a Z boson at $\sqrt{s}$ = 13 TeV with the ATLAS detector. Eur. Phys. J. C **81**, 737 (2021). https://doi.org/10.1140/epjc/s10052-021-09439-4
15. CMS Collaboration, JHEP **03**, 056 (2020). https://doi.org/10.1007/JHEP03(2020)056
16. LHC Higgs Cross Section Working Group, Handbook of LHC Higgs Cross Sections: 4. Deciphering the Nature of the Higgs Sector. Tech. rep. (2016). https://doi.org/10.23731/CYRM-2017-002
17. S. Frixione, V. Hirschi, D. Pagani, H.S. Shao, M. Zaro, JHEP **06**, 184 (2015). https://doi.org/10.1007/JHEP06(2015)184
18. A. Broggio, A. Ferroglia, R. Frederix, D. Pagani, B.D. Pecjak, I. Tsinikos, JHEP **08**, 039 (2019). https://doi.org/10.1007/JHEP08(2019)039
19. A. Kulesza, L. Motyka, D. Schwartländer, T. Stebel, V. Theeuwes, Eur. Phys. J. C **80**(5), 428 (2020). https://doi.org/10.1140/epjc/s10052-020-7987-6
20. W. Bernreuther, D. Heisler, Z.G. Si, JHEP **12**, 026 (2015). https://doi.org/10.1007/JHEP12(2015)026
21. A. Brandenburg, Z.G. Si, P. Uwer, Phys. Lett. B **539**, 235 (2002). https://doi.org/10.1016/S0370-2693(02)02098-1
22. ATLAS Collaboration, Phys. Rev. D **90**(11), 112016 (2014). https://doi.org/10.1103/PhysRevD.90.112016
23. J. Alwall, R. Frederix, S. Frixione, V. Hirschi, F. Maltoni, O. Mattelaer, H.S. Shao, T. Stelzer, P. Torrielli, M. Zaro, JHEP **07**, 079 (2014). https://doi.org/10.1007/JHEP07(2014)079
24. T. Sjöstrand, S. Ask, J.R. Christiansen, R. Corke, N. Desai, P. Ilten, S. Mrenna, S. Prestel, C.O. Rasmussen, P.Z. Skands, Comput. Phys. Commun. **191**, 159 (2015). https://doi.org/10.1016/j.cpc.2015.01.024
25. R.D. Ball et al., JHEP **04**, 040 (2015). https://doi.org/10.1007/JHEP04(2015)040
26. S. Frixione, E. Laenen, P. Motylinski, B.R. Webber, JHEP **04**, 081 (2007). https://doi.org/10.1088/1126-6708/2007/04/081
27. P. Artoisenet, R. Frederix, O. Mattelaer, R. Rietkerk, JHEP **03**, 015 (2013). https://doi.org/10.1007/JHEP03(2013)015
28. N. Castro, J. Erdmann, C. Grunwald, K. Kröninger, N.A. Rosien, Eur. Phys. J. C **76**(8), 432 (2016). https://doi.org/10.1140/epjc/s10052-016-4280-9
29. O. Schulz, F. Beaujean, A. Caldwell, C. Grunwald, V. Hafych, K. Kröninger, S.L. Cagnina, L. Röhrig, L. Shtembari, S.N. Comput, Sci. **2**(3), 210 (2021). https://doi.org/10.1007/s42979-021-00626-4
30. H. Jeffreys, *The Theory of Probability. Oxford Classic Texts in the Physical Sciences (OUP)* (Oxford, 1998). https://books.google.fr/books?id=vh9Act9rtzQC
31. I. Brivio, M. Trott, Phys. Rept. **793**, 1 (2019). https://doi.org/10.1016/j.physrep.2018.11.002
32. O. Bessidskaia Bylund, F. Maltoni, I. Tsinikos, E. Vryonidou, C. Zhang, JHEP **05**, 052 (2016). https://doi.org/10.1007/JHEP05(2016)052
33. B. Grzadkowski, M. Iskrzynski, M. Misiak, J. Rosiek, JHEP **10**, 085 (2010). https://doi.org/10.1007/JHEP10(2010)085
34. G. Durieux, F. Maltoni, C. Zhang, Phys. Rev. D **91**(7), 074017 (2015). https://doi.org/10.1103/PhysRevD.91.074017
35. ATLAS Collaboration, Phys. Rev. D **99**(7), 072009 (2019). https://doi.org/10.1103/PhysRevD.99.072009
36. CMS Collaboration, JHEP **01**, 096 (2016). https://doi.org/10.1007/JHEP01(2016)096
37. CMS Collaboration, JHEP **03**, 095 (2021). https://doi.org/10.1007/JHEP03(2021)095
38. R. Röntsch, M. Schulze, JHEP **07**, 091 (2014). https://doi.org/10.1007/JHEP09(2015)132 **[Erratum: JHEP 09, 132 (2015)]**
39. R. Röntsch, M. Schulze, JHEP **08**, 044 (2015). https://doi.org/10.1007/JHEP08(2015)044
40. M. Schulze, Y. Soreq, Eur. Phys. J. C **76**(8), 466 (2016). https://doi.org/10.1140/epjc/s10052-016-4263-x
41. J.A. Aguilar-Saavedra, Nucl. Phys. B **812**, 181 (2009). https://doi.org/10.1016/j.nuclphysb.2008.12.012
42. D. Barducci et al., Interpreting top-quark LHC measurements in the standard-model effective field theory. Tech. rep. (2018)
43. O. Mattelaer, Eur. Phys. J. C **76**(12), 674 (2016). https://doi.org/10.1140/epjc/s10052-016-4533-7
44. DELPHI Collaboration, Eur. Phys. J. C **60**, 1 (2009). https://doi.org/10.1140/epjc/s10052-009-0917-2
45. N.P. Hartland, F. Maltoni, E.R. Nocera, J. Rojo, E. Slade, E. Vryonidou, C. Zhang, JHEP **04**, 100 (2019). https://doi.org/10.1007/JHEP04(2019)100
46. J. Brehmer, K. Cranmer, F. Kling, T. Plehn, Phys. Rev. D **95**(7), 073002 (2017). https://doi.org/10.1103/PhysRevD.95.073002